\documentstyle[graphics,psfig]{aa}
\newcommand{\ls}
 {\mathrel{\hbox{\rlap{\hbox{\lower4pt\hbox{$\sim$}}}\hbox{$<$}}}}
\newcommand{\gs}
 {\mathrel{\hbox{\rlap{\hbox{\lower4pt\hbox{$\sim$}}}\hbox{$>$}}}}

\def\la{\mathrel{\hbox{\rlap{\hbox{\lower4pt\hbox{$\sim$}}}{\raise2pt\hbox{$<$}}}}}
\def\ga{\mathrel{\hbox{\rlap{\hbox{\lower4pt\hbox{$\sim$}}}{\raise2pt\hbox{$>$}}}}}

\begin{document}

\title{XMM-Newton observation of the Seyfert 1 ESO\,198-G24}
\author{D. Porquet\inst{1} 
\and J.S. Kaastra\inst{2}
\and K.L. Page\inst{3} 
\and P.T. O'Brien\inst{3}
\and M.J. Ward\inst{3}
\and J. Dubau\inst{4}
}

\offprints{Delphine Porquet,\\
 \email{dporquet@mpe.mpg.de}}

\institute{Max-Planck-Institut f{\"u}r extraterrestrische Physik, Postfach 1312, D-85741 Garching, Germany
\and SRON, Sorbonnelaan 23584 CA Utrecht, The Netherlands
\and X-Ray Astronomy Group; Department of Physics and Astronomy; Leicester University; LE1 7RH; U.K.
\and LIXAM, Universit{\'e} Paris-sud, F-91405 Orsay Cedex, France
}
\date{Received ...  / accepted ...}

\abstract{
We present the results from an {\sl XMM-Newton} observation (January 24, 2001) 
of the bright Seyfert 1  galaxy \object{ESO 198-G24} (z=0.045). 
 We found that this Seyfert has an intrinsic 2--10\,keV luminosity of about 10$^{44}$ erg\,s$^{-1}$. 
This source shows no intrinsic absorption in addition to
the Galactic absorption  (${\cal N}_{\rm H}\sim$3$\times$10$^{20}$\,cm$^{-2}$). 
We found both with EPIC and RGS 
that this source possesses significantly steeper 
spectra below $\sim$1.5--2\,keV than observed at higher X-ray energies, 
the so-called soft X-ray excess.   
 The RGS spectra reveal no significant narrow absorption lines 
 suggesting that  if there is a warm absorber, 
it either has a relatively low column density, 
or a very high ionization  parameter.  
 The RGS data are well described by the combination 
 of  a power-law,  a modified black body continuum,
 and weak relativistic lines of  \ion{O}{viii}, and \ion{C}{vi} Ly$_{\alpha}$. 
 However other interpretations are not definitely excluded. 
 The 2--10\,keV energy band  is well fitted by an absorbed power-law 
with a photon spectral index of $\Gamma$=1.77$\pm$0.04 (consistent  
with the typical $\Gamma \sim$1.7 found in Seyfert\,1 galaxies). 
 We found the presence of a narrow Gaussian emission line
  at 6.41\,keV  (i.e $<$ \ion{Fe}{xvii}) with a  moderate equivalent width of about 60--70\,eV, 
and we found an upper limit for
a broad component, if any, of 75\,eV.
We also found a weak absorption edge associated with cold iron with an optical depth of about 0.2.
 
\keywords{galaxies: active -- quasars: individual: ESO\,198-G24 -- X-rays: galaxies} 
}
\maketitle
\titlerunning{ESO\,198-G24 observed with XMM-Newton.}
\authorrunning{Porquet et al.}

\section{Introduction}

The Seyfert 1 ESO\,198-G24  (z=0.045) is one object of the Piccinotti 
Sample of Active Galactic Nuclei (Piccinotti et al. \cite{Piccinotti82}). It was already observed 
in the soft X-ray band by the {\sl ROSAT PSPC} (Turner et al. \cite{Turner93}) 
during two distinct observations (December 1991 and July 1992). 
This source was selected because of a low absorbing column density 
(${\cal N}_{\rm H}\sim$ 3 $\times$ 10$^{20}$cm$^{-2}$) along the 
line-of-sight. It was found that this source possesses a significantly 
steeper spectrum below $\sim$1\,keV than observed at higher X-ray energies. 
During the {\sl ROSAT} December observation, 
the soft X-ray spectrum of ESO\,198-G24 suggested a 
spectral emission or absorption feature, either a Gaussian emission line 
 at energy 0.75\,keV$\pm$0.04\,keV with an equivalent width (EW) of 99$\pm$30\,eV
 (most readily identified as a blend of emission species dominated by ionized iron and oxygen),
 or an absorption edge at E=1.16~$\pm$0.06\,keV ($\tau$=0.37$\pm$0.16).
On the contrary during the {\sl ROSAT} July observation, the spectrum with a 
similar signal-to-noise ratio was adequately described by a simple featureless absorbed 
power-law model. 
Turner et al. (\cite{Turner93}) noted that the spectral feature appeared in the brighter  
state observation of ESO\,198-G24, as might be expected if it originates 
in material responding to ionization by the active nucleus. 
The only constraint on the variability time scale was that it must be 
less than 6 months. 
ESO\,198-G24 was observed  at higher energies 
by {\sl BATSE} on board {\sl CGRO} (Malizia et al. \cite{Malizia99}), and
was reported for the first time as a hard X-ray emitting source: 
F$_{_{\small {\rm MEAN}}}$(2--100\,keV) = 5.27 $\times$ 10$^{-11}$\,erg\,cm$^{-2}$\,s$^{-1}$ 
(weighted mean calculated over nearly 4 years of observations).
 In the 2--10\,keV energy band a variability factor of  1.31 was observed,  
with a mean unabsorbed flux of 3.3 $\times$ 10$^{-11}$\,erg\,cm$^{-2}$\,s$^{-1}$.\\
\indent Very recently, Guainazzi (\cite{Guainazzi2003}) presented results 
  from {\sl ASCA} (July 8$^{\rm th}$, 1997; $\sim$40\,ks), 
{\sl XMM-Newton} (December 1$^{\rm rst}$, 2000; $\sim$9\,ks, guaranteed time),
 and {\sl BeppoSAX} (January 23$^{\rm th}$, 2001; $\sim$150\,ks) data.
He focused on the study of the Fe\,K$_{\alpha}$ line, and 
found that the intensity and line profile are different between
 these three observations. During the {\sl ASCA} observation 
the line profile at 6.4\,keV is narrow without any indication 
of a broad component. On the contrary the {\sl XMM-Newton} PN spectrum 
  (effective time duration of about 6.8\,ks)
 displayed a broader line centered at 6.4\,keV with an additional weak feature at 5.7\,keV.
Guainazzi suggested that this may be one example of ``double-horned profile'', 
 similarly to the Seyfert\,1 \object{MCG-6-30-15} 
 (e.g., Tanaka et al. \cite{Tanaka95}, Wilms et al. \cite{Wilms2001}).

We present here the observation of ESO\,198-G24 in January 24$^{\rm th}$, 2001 
obtained with the X-ray satellite {\sl XMM-Newton}. This present work is the first analysis 
 for this object combining moderate (EPIC) and high resolution data (RGS). 
  Section~2 details the observation and data reduction procedures. 
 Section 3 reports the spectral analysis of the EPIC data 
 (2--5\,keV continuum, the Fe\,K$_{\alpha}$ line region, 
and the broad band 0.3--10\,keV energy band)  and of the RGS data.
 The results are discussed in section 4. 

\section{XMM-Newton observation and data analysis}

ESO\,198-G24 was observed by {\sl XMM-Newton} on January 24, 2001  
 (AO-1, $\sim$30~ks duration). 
The observations with the EPIC MOS (Turner et al. \cite{Turner2001}), 
PN (Str{\"u}der et al. \cite{Strueder2001}) 
and the RGS (den Herder et al. \cite{denHerder2001}) detectors were done.
There are no other strong X-ray sources seen within
the field of view. The MOS and PN cameras (EPIC) operated 
in standard Full Frame mode using the medium filter.  
The data are screened and re-processed with the XMM SAS 
(Science Analysis Software), version 5.3.3. 
The EPIC data cleaning was improved by rejecting solar flares.
After this data cleaning, we obtain as net exposure times about 22.9\,ks, 
23.1\,ks, and 15.6\,ks for MOS1, MOS2, and PN respectively.
 Only X-ray events corresponding to single events 
(i.e. pattern 0) are used  for both MOS and PN, 
in order to reach the best spectral resolution, 
and limit effect of pile-up.
  We note for further comparison that Guainazzi (\cite{Guainazzi2003}) 
used single and double events for spectral analysis of the 
December 2000 {\sl XMM-Newton} observation.
  There was negligible flux variability of
ESO\,198-G24 ($<$10$\%$) during the present observation, therefore we
use the cumulative data for all our subsequent analyses. 
 We then proceed to extract spectra for both source
and background for each EPIC detector. 
A circular source region is defined around the
centroid position of ESO\,198-G24, 
of 1\arcmin\, radius for the MOS, and of 40\arcsec\, for the PN (to avoid the edge
of the chip). The majority of source counts fall onto the source regions 
(at least 95$\%$).
Background spectra are taken from an annular radius centered on ESO\,198-G24, 
between 3.2\arcmin~ and 5.2\arcmin~ (excluding X-ray point sources). 
The background subtracted spectra are fitted, using {\sc xspec} v11.2.0.
 We use the following response matrices:
 {\small m1$\_$medv9q19t5r5$\_$p0$\_$15.rsp, 
m2$\_$medv9q19t5r5$\_$p0$\_$15.rsp,
 epn$\_$ff20$\_$sY9$\_$medium.rsp}.  
The spectra are binned to a minimum of 20 counts per bin  
to apply the $\chi^2$ minimalisation technique.\\
\indent  Both RGS cameras are operated in the standard spectroscopy + Q 
mode.  The RGS data are improved by rejecting solar flares.
After this data cleaning, we obtain as net exposure time about 24.2\,ks 
and 23.6\,ks for respectively RGS1 and RGS2. \\

Further on, values of H$_{0}$=50\,km\,s$^{-1}$\,Mpc$^{-1}$ 
and q$_{0}$=0 are assumed. 
All the fit parameters are given in the galaxy rest-frame.
The errors quoted correspond to 90$\%$ 
confidence ranges for one interesting parameter 
($\Delta \chi^{2}$=2.71).
 The cross-sections for X-ray absorption by the interstellar medium used
 throughout this work  are from Wilms et al. (\cite{Wilms2000}),
 and the element abundances are from Anders \& Grevesse (\cite{AG89}). 

\section{Spectral analysis}

\subsection{The EPIC data}
At a first step, we fit the EPIC data with an absorbed power-law over the 0.3--10\,keV bandpass. 
We do not obtain an acceptable fit with a single absorbed power-law: 
$\Gamma$= 1.92$\pm$0.01 ($\chi^{2}$/d.o.f.= 1825.341/1537,
 with d.o.f. degrees of freedom). 
We do not find any additional intrinsic absorbing medium to the 
 Galactic column density, i.e. 3.09$\times$10$^{20}$\,cm$^{-2}$ 
 (obtained with the tool {\sc coldens}\footnote{http://asc.harvard.edu/toolkit/colden.jsp}). 
 Therefore further on, we fix the column density to the above Galactic value. 
 We split our following spectral analysis in several energy bands: 
 2--5\,keV, the Fe\,K$_{\alpha}$ line region, 
the 0.3--10\,keV broad band energy including the soft excess feature. 

\subsubsection{The 2--5 keV energy band}

\begin{figure}[!ht]
\psfig{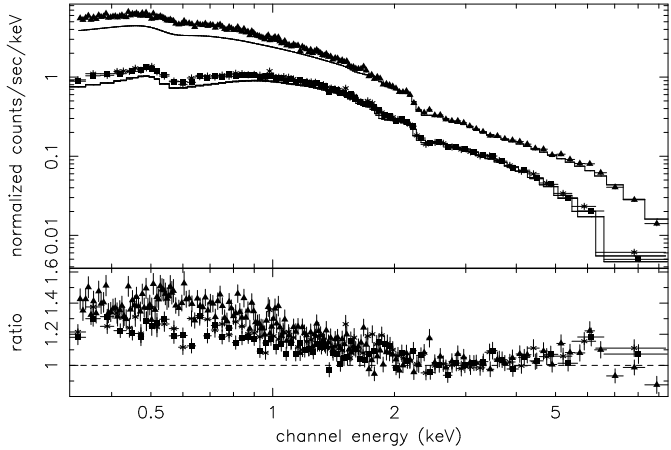}
\caption{The EPIC spectra of ESO\,198-G24 (observer frame and binning at 20\,$\sigma$). 
An absorbed power-law ($\Gamma\sim$1.77) has been fitted within the 2--5\,keV energy range, 
and extrapolated to lower and higher energies. 
A broad  soft X-ray excess is clearly 
seen extending to almost 1.5--2\,keV,
 as well a deviation near 6.2\,keV. 
{\it Filled squares}: MOS1, {\it stars}: MOS2, and {\it triangles}: PN.} 
\label{fig:fig1}
\end{figure}

We begin studying the spectra by fitting an 
absorbed power-law model 
to the EPIC spectra in the 2--5\,keV energy band,
 where the spectrum should be relatively unaffected by possible presence
 of a soft excess or a Warm Absorber-Emitter  
 medium, of  Fe\,K$_{\alpha}$ line emission, 
and of a contribution above 8\,keV that 
could be associated with a high energy reflection hump.
The data are very well fitted by a single power-law model
 with  $\Gamma$= 1.77$\pm0.04$  
($\chi^{2}$/d.o.f.= 681.3/713).  
This photon index value is entirely consistent with 
 those found by Guainazzi (\cite{Guainazzi2003}) 
 for different time observations (in order of increasing 2--10\,keV flux):
 1.75$^{+0.05}_{-0.03}$ ({\sl ASCA}, July 1997), 
 1.77$\pm$0.03 ({\sl XMM-Newton}, December 2000), 
 and   1.79$\pm$0.04 ({\sl BeppoSAX}, January 2001). 
Figure~\ref{fig:fig1} displays the data to model ratio 
 extrapolated over the  0.3--10\,keV broad band energy. 
A significant positive residual is clearly seen below 1.5--2\,keV, 
both in MOS and PN data. 
This could be due to the presence of a soft excess due for example to either 
emission from a cold/ionized accretion disk, 
or complex absorption and/or emission from 
highly ionized gas, the so-called ``Warm Absorber'' (WA). 
There is also a positive deviation near 6.2\,keV (in the observer frame), 
suggesting the presence of an Fe\,K$_{\alpha}$ complex emission line.

\subsubsection{The Fe\,K$_{\alpha}$ line}

The Fe\,K$_{\alpha}$ emission line near 6.4\,keV was found to be a common feature 
 in the hard X-ray spectra of many broad-line Seyfert 1 galaxies 
 (e.g., Pounds et al. \cite{Pounds90}, Nandra \& Pounds \cite{Nandra94}).
 It has been recognized as a powerful  diagnostic of the 
innermost regions of AGN. 

\begin{figure}
\psfig{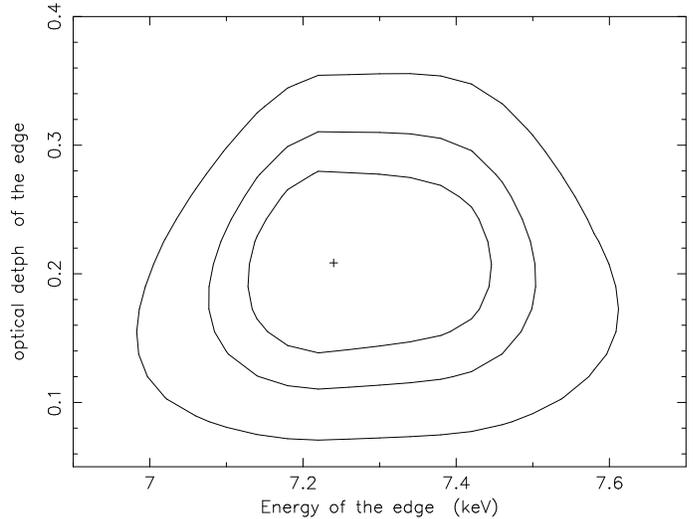}
\caption{The contour plot of the optical depth of the 
 absorption edge versus its energy, 
 for $\Delta\chi^{2}$= 2.3, 4.61, and 9.21  
 (68$\%$, 90$\%$, and 99$\%$ confidence levels, respectively).}
\label{fig:Eedge-tau}
\end{figure}

\begin{figure}
\psfig{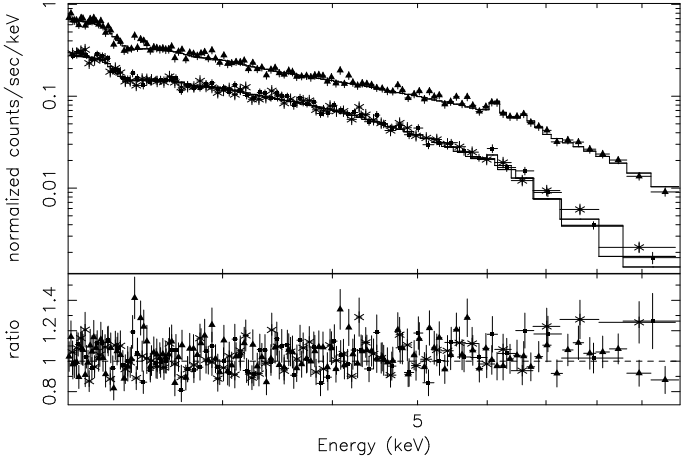}
\caption{The 2--10\,keV EPIC spectrum of ESO\,198-G24 
(observer frame and binning of 10\,$\sigma$).
 An absorbed power-law, two Gaussian emission lines ($\sigma$=10\,eV) 
and one absorption edge have been fitted.
 {\it Filled squares}: MOS1, {\it stars}: MOS2, and {\it triangles}: PN.}
\label{fig2}
\end{figure}

In the 2--10\,keV energy band, a simple absorbed power-law model 
 gives a good fit (see Table~\ref{table:FeK}), with an unabsorbed 2--10\,keV 
 flux of 1.10$\pm$0.04$\times$10$^{-11}$\,erg\,cm$^{-2}$\,s$^{-1}$
 (i.e., L$_{\rm x}$(2--10\,keV)$\sim$10$^{44}$\,erg\,s$^{-1}$). 
  It is consistent with the value found during the 
 previous {\sl XMM-Newton} observation in December 2000 
(i.e., 1.09$\pm$0.01 $\times$10$^{-11}$ erg\,cm$^{-2}$\,s$^{-1}$, Guainazzi \cite{Guainazzi2003}). 
 There is still a clear positive deviation near 6.4\,keV (in the rest-frame), 
  with this power-law model with $\Gamma\sim$1.77 
 (which is the same index value as found in the 2--5\,keV energy range, see Figure~\ref{fig:fig1}).  
 The addition of a narrow  (i.e., the line width $\sigma$ is fixed to 10\,eV, 
  i.e. intrinsically narrow) Gaussian emission 
 line improves significantly the adjustment of the data 
($\Delta \chi^{2}$=30 for only two additional parameters, 
 F-test probability of 99.99$\%$, see Table~\ref{table:FeK}). 
The rest-frame energy of the line at about  6.41\,keV 
 corresponds to  fluorescence from iron 
in a ``low'' state of ionization (i.e., $\leq$ \ion{Fe}{xvii})
 and its equivalent width (EW) is consistent with the typical values measured 
in Seyfert\,1 galaxies ($\sim$ 50--150\,eV; 
e.g., Nandra \& Pounds \cite{Nandra94}). 
We let free the line width, but we do not obtain strong 
constraint on this parameter with $\sigma<$112\,eV (Table~\ref{table:FeK}). 
 Now we look for a possible weak broad line component, 
 then we test a relativistic line model (Laor \cite{Laor91}).
We fix the parameters of the line to those inferred from the RGS data 
for low-Z ions \ion{O}{viii}, and \ion{C}{vi} Ly$_{\alpha}$ (see $\S$\ref{sec:rgs}).
 The line is badly fitted with only this relativistic line profile, then 
we add to the {\sc laor} profile a narrow line component ($\sigma$=10\,eV).
   We obtain an upper limit for the EW of the {\sc laor} profile of 75\,eV, 
and EW of 61$^{+21}_{-27}$\,eV for the narrow component.
 However the $\Delta\chi^{2}$ is less than unity for one additionnal parameter, 
 comparing to the model with the narrow component line alone,  see Table~\ref{table:FeK}, 
  therefore  there is no statistical requirement 
for a broad component for the Fe\,K$_{\alpha}$ line 
 at 6.4\,keV in the present observation.  
Similar results are obtained when using the relativistic line parameters 
at 6.4\,keV reported in Table 2 in Guainazzi (\cite{Guainazzi2003}). 
We also test the possible presence of a line component at 5.7 keV
claimed by Guainazzi (2003) in the December 2000 XMM-Newton
observation. Analysing the data of December 
 2000, we do not find any statistical requirement for a line at 5.7\,keV 
(rest-frame) with EW$<$37\,eV. 
As well this component is not required in the present January 2001 data
 with EW $<$ 23 eV  and an F-test probability 
 of only 40$\%$ (Table~\ref{table:FeK}).  
The addition of a second narrow Gaussian emission line at higher energy (E$>$6.4\,keV) 
 yields a further very slight improvement  
 ($\Delta \chi^{2}\sim$5  for 2 additional parameters,  F-test probability of 94.8$\%$, 
 see Table~\ref{table:FeK}). 
This second emission line is less well determined but it clearly lies 
to the high-energy side of the ``cold'' line with a best-fit energy 
 E=6.92\,keV (most probably due to \ion{Fe}{xxvi}, i.e H-like iron). 
 There is a negative deviation compared 
to the model near 7\,keV (in the galaxy frame), 
therefore we add an absorption edge and we obtain a better fit
 ($\Delta\chi^{2}$=16 for 2 additional parameters, 
  F-test probability of 99.97$\%$, 
 Table~\ref{table:FeK}). 
Its energy is consistent with absorption due to cold to moderate 
ionized iron ions (i.e., $<$ \ion{Fe}{xii}). 
 Figure~\ref{fig:Eedge-tau} displays the contour plot for the optical depth 
 of the edge versus its energy, 
 for $\Delta\chi^{2}$=2.3, 4.61, and 9.21 
(68$\%$, 90$\%$, and 99$\%$ confidence levels, respectively). 
The ionization state of the ``cold'' emission line and the absorption edge are compatible.\\
\indent  To summarize the present data are well represented 
 in the 2--10\,keV energy range by a power-law 
 and by a narrow Fe\,K$_{\alpha}$ line 
 at about 6.4\,keV, and a weak absorption edge near 7.2\,keV. 
 Both energies are compatible with iron nearly or moderatly ionized (i.e., $<$\ion{Fe}{xvii}
 and $<$\ion{Fe}{xii}, respectively).
 Also there is a slight indication for a very ionized narrow Fe\,K$_{\alpha}$ emission 
line near 6.9\,keV, which most probably corresponds to \ion{Fe}{xxvi}, i.e H-like iron.
 See Table~\ref{table:FeK} and Figure~\ref{fig2}.\\ 

\begin{table*}[!ht]
\caption{Fit of the EPIC data in the 2--10\,keV energy range. 
$E$ is the energy (in keV) of the corresponding emission line or
absorption edge. $EW$ is the equivalent with of the emission line in eV. 
 $\tau$ is the optical depth of the absorption edge. 
 F-test is the F-test probability adding one component compared to the single power-law model. 
}
\begin{tabular}{cccccccc}
\hline
\hline
\noalign {\smallskip}
 Model                                &   $\Gamma$    & E$_{\rm line}$ or E$_{\rm edge}$&  $\sigma_{\rm line}$&  F$_{\rm line}$  &  EW$_{\rm line}$  (eV)     &    F-test   &   $\chi^{2}$/d.o.f.   \\
                                           &                        &  (in keV)                                      &       (eV)                              &     {\small (erg\,cm$^{-2}$\,s$^{-1}$)}  &  $\tau_{\rm edge}$              &            &                              \\
                                             \\ 
\noalign {\smallskip}
\hline
\noalign {\smallskip}
{\small PL}           & $\Gamma$=1.77$\pm$0.03            &   $-$                                          &     $-$               &  $-$            &                      $-$                             &           $-$                      &{\small 986.0/973} \\
\noalign {\smallskip}
\hline
\noalign {\smallskip}
{\small PL+ GA}           &  $\Gamma$=1.78$\pm$0.03   &6.41$^{+0.02}_{-0.04}$ & 10 (fixed) &  9.5$\pm2.9\times$10$^{-6}$  &   74$\pm$23 &     99.99$\%$                        &{\small 956.1/971} \\
\noalign {\smallskip}
\hline
\noalign {\smallskip}
{\small PL+ GA}                       &  $\Gamma$=1.78$\pm$0.03    &6.40$\pm$0.04   &   57$^{+55}_{-57}$    &   1.11$^{+0.43}_{-0.39}\times$10$^{-5}$ &   86$^{+33}_{-30}$ &   99.99$\%$               &{\small 955.1/970} \\
\noalign {\smallskip}
\hline
\noalign {\smallskip}
{\small PL+ GA}       & $\Gamma$=1.77$\pm$0.03 &   5.7 (fixed) &   10  (fixed)    & $<$1.2$\times$10$^{-5}$    &  $<$23 &   40$\%$                          &{\small 985.7/972} \\
\noalign {\smallskip}
\hline
\noalign {\smallskip}
{\small PL + GA}                &  $\Gamma$=1.77$\pm$0.03  &  6.92$\pm$0.08 &   10  (fixed)   & 4.2$\pm$2.8$\times$10$^{-6}$ & 37$\pm$25 &  94.8$\%$   &{\small 980.0/971} \\
\noalign {\smallskip}
\hline
\noalign {\smallskip}
{\small PL+ edge}       & $\Gamma$=1.72$\pm$0.03 &    7.25$^{+0.21}_{-0.13}$ &         $-$                       &               $-$                                  &    0.21$\pm$0.08         &   99.97$\%$                &{\small 964.3/971} \\
\noalign {\smallskip}
\hline
\noalign {\smallskip}
{\small PL + 2$\times$GA}   &  $\Gamma$=1.75$\pm$0.03  &  6.41$\pm$0.03 &   10  (fixed)   & 8.8$\pm$2.9$\times$10$^{-6}$ & 68$\pm$22 &   $-$       &{\small 935.7/967} \\
\noalign {\smallskip}
                                                &                                                      &  6.92$\pm$0.10 &   10  (fixed)   & 3.6$\pm$2.7$\times$10$^{-6}$ & 37$\pm$25 &         &                       \\
\noalign {\smallskip}
          + edge    &                      &    7.24$^{+0.25}_{-0.14}$ &         $-$                       &               $-$                                  &    0.17$\pm$0.07         &                    & \\
\noalign {\smallskip}
\hline
\hline
\end{tabular}
\label{table:FeK}
\end{table*}

 The narrow component found here at 6.4\,keV 
is now emerging as a common feature of Seyfert 1 spectra seen
 by {\sl XMM-Newton} and {\sl Chandra}
 (see, e.g., Mrk\,205: Reeves et al. \cite{Reeves2001}; 
NGC\,5548: Yaqoob et al. \cite{Yaqoob2001}; 
\object{NGC\,3783}: Kaspi et al. \cite{Kaspi2001}). 
Detection of a significant narrow iron emission line at 6.4 keV implies 
that a substantial quantity of cool reprocessing material is present. 
The line strength (EW$\sim$ 70\,eV), implies that the cool matter if seen by reflection 
subtends a substantial solid angle, of at least 1  steradian, assuming it is 
Thomson thick and has a solar abundance of iron
(e.g. George \& Fabian \cite{George91}). The most likely location of such material, 
 according to the current AGN models, would appear to be the molecular torus
(Antonucci \cite{A93}), with hard X-rays from the central engine being 
reflected off the inner surface of the torus and into the line-of-sight of the observer 
(Krolik  et al. \cite{Krolik94}, Ghisellini et al. \cite{Ghisellini94}). 
Although reflection off the putative torus is attractive, we cannot rule out other options. 
Emission from BLR clouds could account for some of the narrow line flux as probably found in
\object{NGC 5548} (Yaqoob et al. \cite{Yaqoob2001}), 
although the predicted line strength is small (EW$<$50 eV; Leahy \& Creighton \cite{Leahy93}), 
 but consistent within the error bars with the value found here.  
Another possibility is that the line originates by X-ray reflection from 
the outer regions of the accretion disc. 
 Normally the angle subtended by the outer
disk, to the X-ray source, would be small. However a warped, concave disk could 
increase the amount of reflection at large radii, and account for a measurable narrow
line component at 6.4 keV (Blackman \cite{Blackman99}). 
The present 6.4\,keV emission line is much weaker 
than the one inferred from the  {\sl ASCA} observation 
in July 1997 (i.e. EW=320$^{+70}_{-100}$\,eV for an unresolved Gaussian line). 
 As well, it  also  appears approximately two times weaker 
than the one measured during December 2000, by Guainazzi et al. (\cite{Guainazzi2003}) 
with similar intrinsic 2--10\,keV fluxes.  
However the claim of a possible variability of this line 
 between these two observations should be taken into caution, 
 indeed analysing the December 2000 data 
(assuming a narrow line at 6.41\,keV and a line width of 10\,eV),
 we obtain EW=70$\pm$45\,eV.
 This value is compatible with the one reported in Table 1
 for the present data.
 No reliable comparison is possible with the BeppoSAX data, 
since its spectral resolution 
 is not sufficient to determine whether the profile was narrow or 
 broad, however the EW are compatible within the error bars (i.e. EW=100$^{+30}_{-60}$\,eV).\\ 
 \indent  The emission line near 6.92\,keV, 
if any, corresponding to \ion{Fe}{xxvi} (H-like iron), 
 could be produced by the disc matter in a high state of ionization, as observed 
in  several radio-quiet quasars (Reeves \& Turner \cite{Reeves2000}). 
  As we will show below, the EPIC data may be explained by an ionized accretion disk 
model from  Ross \& Fabian (\cite{Ross93}) and Ballantyne et al. (\cite{Bal2001}). 
Another possibility is the ionized lines originate from the WA, 
which can produce substantial emission from He and H-like iron 
(Krolik \& Kallman \cite{Krolik87}). 
As we will see below, the WA, if present, does 
not show significant absorption or emission features below 2.5\,keV. 
This may mean that either there is only almost no warm absorber material on
the line-of-sight, or the low-Z ions (e.g., O, Ne, Mg, Si) are completely ionized, 
which is consistent with the very high ionized state (H-like) 
found for the Fe\,K$_{\alpha}$ line at about 6.9\,keV.  \\
\indent  As reported by Guainazzi (\cite{Guainazzi2003}), 
no absorption edges existed in the EPIC/PN spectrum 
during the observation of December 2000, 
with only an upper limit of 0.14. 
However, the statistics of this observation is much lower 
than the one presented here, then enable any reliable comparison 
for a possible variability of the absorption edge. 
 We notice that the absorption edge found in the EPIC data near 7.24\,keV 
 (i.e. $<$\ion{Fe}{xii}) cannot be produced by the WA, if any.
Indeed, the optical depth found for this absorption edge is about 0.1--0.2, 
which means that the WA would be completly opaque below about 2\,keV, 
 while this is not what we demonstrate in section 3.4. 
The WA for log~$\xi\leq$2 corresponds to too weak column density
to produce a significant absorption edge of cold iron ions as observed at 7.24\,keV. 

\subsubsection {The EPIC broad band spectrum}

\begin{table*}[!ht]
\caption{X-ray spectral fits over the 0.3--10\,keV broad band  energy range. 
In all models we add two Gaussian emission lines and one absorption edge.  
 The energies and the width of the lines are fixed to the values found 
in Table~\ref{table:FeK} for the last model. 
 {\sc bb}: black-body component.  
{\sc pl}: power-law component.  
{\sc pexrav}: exponentially cut off power-law 
spectrum reflected from neutral material
(Magdziarz \& Zdziarski \cite{Magdziarz95}),  
the cutoff energy has been fixed to 200\,keV.  
  {\sc compTT}: Comptonization model of soft photons in a hot plasma 
from Titarchuk (\cite{Titarchuk94}), here the soft photon temperature is set to
be the same for the two hot plasmas. 
{\sc iondisk}: reflection spectrum from an ionized slab of an input power-law 
from Ross \& Fabian (\cite{Ross93}) and Ballantyne et al. (\cite{Bal2001}). 
(a): the line fluxes are expressed in 10$^{-6}$\,erg\,cm$^{-2}$\,s$^{1}$.}
\begin{tabular}{ccccc}
\hline
\hline
\noalign {\smallskip}                       
                                  & Continuum parameters                      &   Line/edge parameters                                                        &  $\chi^{2}$/d.o.f.\\
\noalign {\smallskip}                       
\hline
\noalign {\smallskip}                       
 {\sc bb}+{\sc pl}     &   kT$_{\rm bb}$=172$\pm$7\,eV  ~~~ $\Gamma$=1.84$\pm$0.02 & F(6.41\,keV)=9.2$\pm$2.7~$^{(a)}$      ~~~  EW=77$\pm$23\,eV     & 1667.9/1535 \\
 \noalign {\smallskip}                       
                                &                                                                                                                                       & F(6.92\,keV)=4.3$\pm$2.6~$^{(a)}$       ~~~  EW=41$\pm$25\,eV    &$\chi^{2}_{_{\rm red}}$=1.09\\
\noalign {\smallskip}                       
                                &                                                                                                                                      &  $\tau_{\rm edge}$(7.24\,keV)=0.10$\pm$0.07               &\\ 
\noalign {\smallskip}                       
\hline
\noalign {\smallskip}                       
 {\sc 2 $\times$ bb}+{\sc pl}     &   kT$_{\rm bb1}$=131$\pm$8\,eV  ~~~  kT$_{\rm bb2}$=317$\pm$28\,eV  &  F(6.41\,keV)= 7.5$\pm$2.7~$^{(a)}$     ~~~  EW=59$\pm$22\,eV     & 1649.8/1536\\
 \noalign {\smallskip} 
                                &          $\Gamma$=1.69$\pm$0.03                          & F(6.92\,keV)$<$ 5$$~$^{(a)}$    ~~~  EW$<$45\,eV    &$\chi^{2}_{_{\rm red}}$=1.07\\
\noalign {\smallskip} 
                                &                                                                                     &   $\tau_{\rm edge}$(7.24\,keV)= 0.25$\pm$0.09              &    \\ 
\noalign {\smallskip}                       
\hline
 \noalign {\smallskip} 
{\sc bb + pexrav}            &   kT$_{\rm bb1}$=186$^{+9}_{-6}$\,eV                              &      F(6.41\,keV)=7.3$\pm$2.8~$^{(a)}$        ~~~  EW=60$\pm$23\,eV  &    1652.8/1534      \\
\noalign {\smallskip}                       
                                &   $\Gamma$=1.88$\pm$0.03 ~~~ R=1.3$\pm$0.5                & F(6.92\,keV) $<$ 4.7~$^{(a)}$       ~~~  EW$<$ 43\,eV    &$\chi^{2}_{_{\rm red}}$=1.08\\
\noalign {\smallskip} 
                                &                                                                                                                                      &   $\tau_{\rm edge}$(7.24\,keV)= 0.15$\pm$0.07               &    \\ 
\noalign {\smallskip}                       
\hline
 \noalign {\smallskip} 
{\sc 2 $\times$ compTT}    & kT$^{1}_{\rm photon}$=  kT$^{2}_{\rm photon}$=10\,eV (fixed)   &      F(6.41\,keV)=9.0$\pm$2.7~$^{(a)}$      ~~~  EW=78$\pm$23\,eV     &  1661.0/1532      \\
 \noalign {\smallskip} 
                                & kT$^{1}_{\rm plasma}$=0.21$\pm$0.01\,keV ~~~  $\tau$=28.1$\pm$0.5   & F(6.92\,keV)=4.0$\pm$2.5~$^{(a)}$     ~~~  EW=40$\pm$25\,eV    &$\chi^{2}_{_{\rm red}}$=1.08\\
\noalign {\smallskip} 
                                &    kT$^{2}_{\rm plasma}$= 200\,keV (fixed)   ~~~    $\tau$= 0.13$\pm$0.02     &   $\tau_{\rm edge}$(7.24\,keV)= 0.12$\pm$0.07              &    \\ 
\noalign {\smallskip}                       
\hline
\noalign {\smallskip}                       
{\sc iondisk}    &  log~$\xi$=4.18$\pm$0.06  ~~~   $\Gamma$=1.63$\pm$0.02 ~~~  R$>$1.7      &      F(6.41\,keV)= 7.7$\pm$2.7~$^{(a)}$      ~~~  EW=62$\pm$22\,eV     &    1642.8/1535      \\
 \noalign {\smallskip} 
                                &                                                                                                                               & F(6.92\,keV)=2.7$\pm$2.6 ~$^{(a)}$    ~~~  EW=24$\pm$23\,eV    &$\chi^{2}_{_{\rm red}}$=1.07\\
\noalign {\smallskip} 
                                &                                    &   $\tau_{\rm edge}$(7.24\,keV)= 0.18$\pm$0.08              &    \\ 
\noalign {\smallskip}                       
\hline
\hline
\end{tabular}
\label{table:0.3-10keV}
\end{table*}
Extrapolating the best-fitting 2--5\,keV power-law model back to 0.3\,keV,  
both MOS and PN spectra clearly reveal the presence of a broad soft excess
below about 1.5--2\,keV as shown in Figure~\ref{fig:fig1}.
 This soft X-ray emission seems to be a common feature in Seyfert 1 galaxies
observed up to now with {\sl XMM-Newton} (see Pounds \& Reeves \cite{Pounds2002}).
The shape of this soft excess emission 
is similar to that found in other high-luminosity Seyfert 1 galaxies 
 (see for example  Fig.~1 in Pounds \& Reeves 2002). \\

 Fitting the 0.3--2\,keV EPIC spectra of ESO\,198-G24 
 by a single  absorbed power-law model,  
 we find a photon power-law index 
 $\Gamma$=1.94$\pm$0.02 ($\chi^{2}$/d.o.f.= 592.6/558, $\chi^{2}_{\rm red}$=1.06), 
 much  harder than those measured according to the two {\sl ROSAT} observations, 
i.e $\Gamma$= 2.28$\pm$0.07  ($\chi^{2}_{\rm red}$= 2.52)
and $\Gamma$= 2.19$\pm$0.07 ($\chi^{2}_{\rm red}$= 1.07) 
respectively for  December 1991 and July 1992 
(Turner et al. \cite{Turner93}).
The unabsorbed 0.1-2\,keV X-ray flux  
 is 1.58$\pm$0.01 $\times$ 10$^{-11}$ \,erg\,cm$^{-2}$\,s$^{-1}$.
 We now investigate different models which may explain 
the 0.3--10\,keV broad band EPIC spectra, i.e. both
the soft excess feature and the hard spectrum. 
In each model, we take into account the iron complex found previously,
 we fix the energy of the two emission lines and of the absorption edge 
 to the values found in Table~\ref{table:FeK} for the last model. 
The inferred parameters for all the following models are reported in Table~\ref{table:0.3-10keV}. 
 First, we fit the spectra with  a black-body (soft X-rays) 
and with a power-law component (hard X-rays). 
We obtain  a good fit for this combined model  
(see Table~\ref{table:0.3-10keV}). 
Adding another black-body component improves the spectral fit 
(i.e. $\Delta\chi^{2}$=18 for 2 additional parameters).
  We also test other disk models such as {\sc diskbb} and {\sc diskpn} 
combined with a power-law, we obtain  as well a good representation of the data,
with kT$\sim$250\,eV. 
 The temperatures found here are rather 
large for a standard steady state $\alpha$ thin accretion disc.
Assuming a black hole (BH) mass of about 3$\times$10$^{8}$\,M$_{\odot}$ 
 (Rokaki \& Boisson \cite{RB99}),  we expect a temperature of about only
 10\,eV at 3\,R$_{s}$ (e.g, Peterson \cite{P97}). 
 Therefore a standard steady state $\alpha$ thin accretion disc
  is ruled out to explain the soft excess observed in ESO\,198-G24. \\
 The broad band X-ray spectrum of ESO\,198-G24 may be explained 
by reflection from an accretion disc (e.g., {\sc pexrav}, Magdziarz \& Zdziarski 
\cite{Magdziarz95}). 
 Since the energy range of our spectral fit is restricted, we fix  
  the cut-off energy to 200\,keV. 
This model gives 
a slightly poorer  fit  ($\chi^{2}$/d.o.f.=1701.1/1536, i.e. $\chi^{2}_{_{\rm red}}$=1.11) 
 than the other ones reported in Table~\ref{table:0.3-10keV}. 
Adding an emission component from the accretion disk 
 (e.g., bb), we obtain a better fit with 
($\Delta\chi^{2}$= 48 for 2 additional parameters, see Table~\ref{table:0.3-10keV}).
 We obtain a reflection component close to 1.3,  
 i.e. is consistent with a reflection from a cold material 
 with a 2.6$\pi$ steradian solid angle.  
 However the black-body temperature is too high 
for a standard steady state $\alpha$ thin accretion disc 
 as mentioned above. Therefore the BB+pexrav model  is also ruled out.\\
\indent Comptonization has often been suggested 
as a source of both the soft X-ray and 
hard X-ray spectrum of AGN, as for example 
 the accretion disc may be responsible for the soft emission and part of these 
soft photons are inversed-Compton scattered 
to the hard X-ray energy range, 
 as they pass through the hot corona above the disc.
 We use  a double Comptonization model  
({\sc compTT}: Titarchuk \cite{Titarchuk94})
 as  in O'Brien et al. (\cite{O'Brien2001}) 
in which soft photons from the accretion disc are up-scattered 
by thermal electrons characterized by two temperatures, to yield
the observed broad soft excess as well as the power-law 
shape at higher energies. We fix the temperature of the comptonizing 
 photons at 10\,eV as expected for a BH mass of 3$\times$10$^{8}$\,M$_{\odot}$
 as mentioned above, and the plasma temperature to 200\,keV for the second {\sc compTT}. 
 The data are well fitted with this model (Table~\ref{table:0.3-10keV}). \\
\indent An ionized accretion disk model 
(Ross \& Fabian \cite{Ross93}, Ballantyne et al. \cite{Bal2001})
 provides also a good fit to the data with a very high ionization parameter 
  ($\xi$=L$_{\rm bol}$/ (n$_{\rm H}~r^{2}$), with 
 $L_{\rm bol}$ the bolometric luminosity in erg\,s$^{-1}$, 
 $n_{\rm H}$ the Hydrogen density of the ionized medium in cm$^{-3}$, 
and   $r^{2}$ the distance of the ionized medium from
the photo-ionizing source in cm). 
 The  value of $\xi$ found here is consistent with an emission 
of a H-like Fe\,K$_{\alpha}$ line (e.g., Fig.~6 from Colbert et al. \cite{Colbert2002}), 
 and with the small residual found at 6.92\,keV in this model. 
 The narrowness of this iron H-like line indicates that the line is emitted 
 far away from the BH or that the disk has a relatively small optically depth 
 for Compton scattering.
  However, we note that the iondisk model used here does not take into
account smearing of the emission lines and hence the derived ionisation
parameter may be overestimated.

 Two models can explain the broad band spectrum of 
ESO\,198-G24: a double comptonization models or an ionized accretion disk.
\subsection{The RGS data analysis}\label{sec:rgs}

The RGS spectrum offers us for the first time a view of
ESO\,198-G24 at very high spectral resolution.
While the spectrum has a rather poor S/N ratio, we are able
to test several models using the SPEX code 
(Kaastra et al. \cite{Kaastra2002a}, version 2.0). 

\begin{figure*}[!ht]
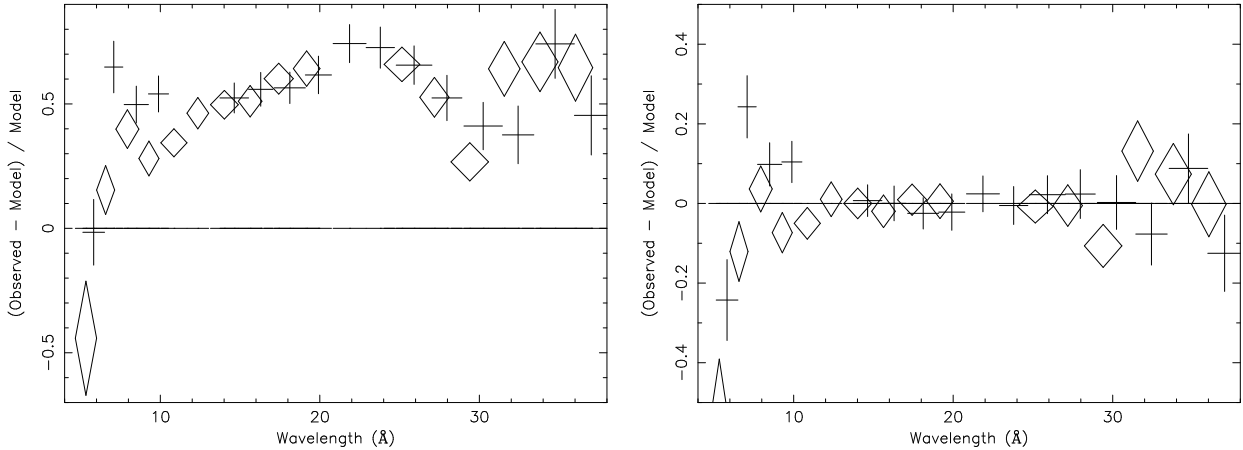

\begin{tabular}{cc}
\psfig{file=ms3632_fig4a.ps,angle=-90,width=8cm} & \psfig{file=ms3632_fig4b.ps,angle=-90,width=8cm} \\
\end{tabular}
\caption{The RGS fit residual spectra of ESO\,198-G24 (observer frame) 
for the power-law (PL) model with the parameter fixed to the values
found with EPIC. 
{\it Left panel}: PL only. 
{\it Right panel}: PL + ``knak'' model (see text for explanations).  
 {\it Crosses}: RGS\,1, {\it diamonds}: RGS\,2.}
\label{fig:RGS}
\end{figure*}

\begin{figure}[!ht]
\psfig{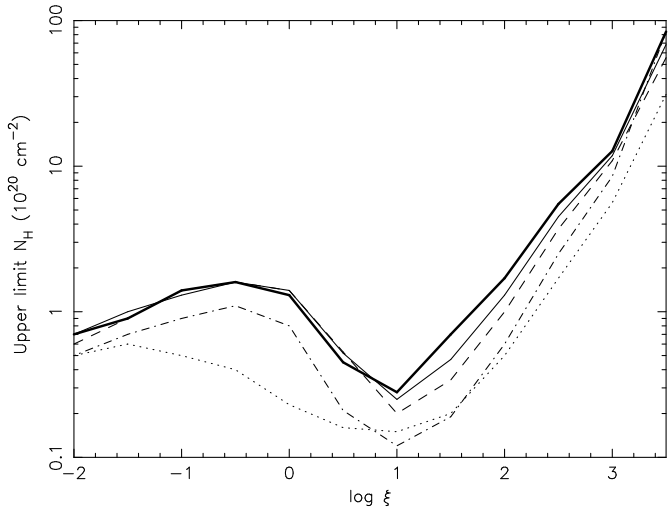}
\caption{Upper limit at 1$\sigma$ of the column density ${\cal N}_{\rm H}$ 
(expressed in 10$^{20}$\,cm$^{-2}$), 
versus the ionization parameter $\xi$ (in logarithm scale). 
The curves correspond to different values of outflow velocities.
{\it Thick solid line}:  $v$=0\,km\,s$^{-1}$, 
{\it thin solid line}: $v$=$-$250\,km\,s$^{-1}$, 
{\it dashed line}: $v$=$-$500\,km\,s$^{-1}$, 
{\it dash-dot line}: $v$=$-$1000\,km\,s$^{-1}$, 
 and {\it dots}: $v$=$-$2000\,km\,s$^{-1}$.}
\label{fig:RGS_nh}
\end{figure}

\begin{figure}[!ht]
\begin{tabular}{cc}
\psfig{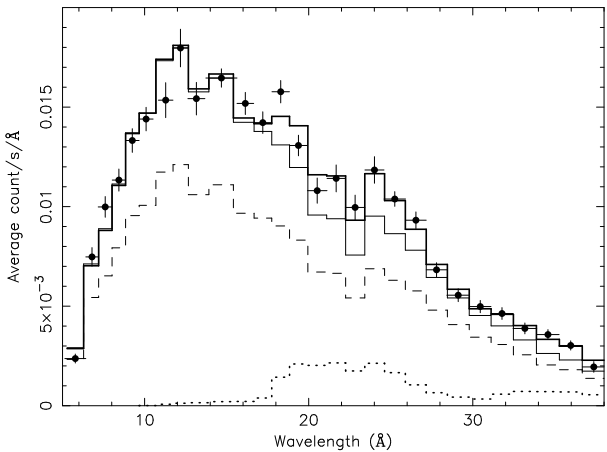} \\
\end{tabular}
\caption{Average RGS spectrum (points), 
with superposed different components: 
 the PL (dashed line), the relativistic {\sc laor} lines (dotted line), 
 PL+{\sc mbb} (thin solid line), 
and the best-fit model (thick solid line):
  PL + {\sc mbb} +  and relativistic {\sc laor} lines for 
\ion{O}{viii}, \ion{N}{vii} and \ion{C}{vi} Ly$_{\alpha}$.  
}
\label{fig:RGSfit}
\end{figure}

First, we start fitting the RGS\,1 and RGS\,2 spectra, 
with the EPIC best-fit parameters of the power-law model. 
The original spectrum was binned to two bins per CCD.
The residuals show a clear soft excess 
(see fig.~\ref{fig:RGS} {\it left panel}). 
 Then we modify the power-law model 
with the  ``knak'' model which is a multiplicative model.  
 It multiplies the spectral model by a factor $f(\lambda)$.  
 The full wavelength range is divided into segments with bounding
wavelengths $\lambda_i$.  The multiplicative scaling factor $f_i$ for each
segment boundary $\lambda_i$ is a free fitting parameter.  For each wavelength
the multiplication factor $f(\lambda)$ is determined by linear interpolation in
the $\log \lambda_i - \log f_i$ plane.  We have chosen the grid points
$\lambda_i$ at 6, 12, 18, 24, 30 and 36~\AA. 
  This empirical model gives a much better fit with $\chi^2$=51.8 for 26 d.o.f. 
 (see fig~\ref{fig:RGS} {\it right panel}). 
Half of the $\chi^2$ value is due to the shortest wavelength bin of each RGS,
 and this is mainly due to remaining calibration uncertainties below  $\sim$7~\AA.
 We obtain that the source as 
for example a soft excess  amplitude of 66$\%$ at 36~\AA, and  44$\%$ at 12~\AA.\\
\indent Evident absorption/emission features are not visible.
Therefore, if there is a WA, 
it either has a relatively low column density, 
or a very high ionization  parameter ($\xi$=$L$/($n_{\rm H}$\,$r^{2}$)).  
The WA is an optically thin ionized medium, 
 detected in most Seyfert 1 galaxies with low to moderate X-ray luminosities 
($L_{X}<$10$^{44}$\,erg\,s$^{-1}$). 
 Up to now, RGS spectra confirm the absence of strong absorption in the high luminosity 
sources (e.g., \object{Mrk 509}: Pounds et al \cite{Pounds2001};  
\object{PKS 0558-504}; O'Brien et al. \cite{O'Brien2001}). 
Low density warm absorbers are most easily detected from their
narrow absorption lines (see for example Kaastra \cite{Kaastra2001}).
To test the presence of warm absorbing gas, we therefore use the above
empirical continuum model (PL + knak) and apply the ``xabs'' WA  
model (see Kaastra et al.  \cite{Kaastra2002b}). The velocity broadening of the 
absorber was fixed to $\sigma_v$ = 100 km \,s$^{-1}$; since we expect low column
density, the equivalent line width does not depend upon $\sigma_v$ but merely upon
the column density.
We use the spectrum with a binning factor of 3.
We search a grid of ionisation parameters (log $\xi$ from -2 to +3.5)
with a set of outflow velocities: $v$= 0, $-$250, $-$500, $-$1000 and $-$2000 km\,s$^{-1}$,
but found no clear detection of a WA. 
The 1\,$\sigma$ upper limits to the
column densities are shown in Figure~\ref{fig:RGS_nh}.
  If any WA is present, 
its column density is low for log~$\xi\leq$2. 
Even in case of a very highly ionised medium  
(i.e., here higher column density upper limit),   
the low-Z ions which absorbed in the soft X-ray range
 are completely ionised, then in both cases 
(moderately or very highly ionised medium)  
 we do not expect strong absorption 
edges. 
We also can put limits on narrow line emission:
the best-fit \ion{O}{vii} $f$ (forbidden) line has  
 a strength of $<$3.2$\times$10$^{50}$ ph\,s$^{-1}$, 
the best-fit \ion{Ne}{ix} $f$ line has a strength 
of 0.4$\pm$0.8$\times$10$^{50}$ ph\,s$^{-1}$
Thus, no significant narrow line emission is found. \\
\indent As a last step, we investigate the broad continuum emission features.  
We find a best-fit model using the combination 
 of a power-law,  a modified black body continuum ({\sc mbb}),
 and weak relativistic {\sc laor} lines for 
\ion{O}{viii},  and \ion{C}{vi} Ly$_{\alpha}$ 
(see Table~\ref{table:bestfit}). 
The power-law parameters are kept fixed to the EPIC values. 
Since there are no narrow features in the spectrum,
we bin the spectrum by a factor of 100 for the spectral fitting. 
Figure~\ref{fig:RGSfit} displays the average RGS spectrum, 
  with superposed the contribution of the power-law,
    the power-law plus modified black body continuum (MBB), and the best fit
    model including all components. 
 The Laor line profiles are intrinsically 
 very broad and asymmetric, and moreover, they
are folded with the instrumental effective area, which also contains
significant broad band structure. The dotted line in  Figure~\ref{fig:RGSfit}  
 displays the sum of the \ion{C}{vi} and \ion{O}{viii}  contributions, and
due to the cosmological redshift, the line spectrum is shifted by about
 5$\%$ or 1--2\,\AA~ for the relevant part.
The bright bin around 24\AA~ is the red wing of the \ion{O}{viii}  line
 (not the peak of a \ion{N}{vii} line), the part
between 28 and 32\,\AA~ is a combination of the reddest part 
of \ion{O}{viii} line with the blue wing of the \ion{C}{vi} line, 
 and the part above 32\,\AA~ is dominated by the red wing of the \ion{C}{vi} line.

\begin{table}
\caption{Best-fit parameters for the average RGS spectrum 
with a model combining a power-law, 
a modified black body continuum ({\sc mbb}),
 and relativistic {\sc laor} lines for 
\ion{O}{viii}, \ion{N}{vii} and \ion{C}{vi} Ly$_{\alpha}$. 
The power-law parameters are kept fixed to the EPIC values. }
\begin{tabular}{cc}
\hline
\hline
\noalign {\smallskip}
$\chi^{2}$/d.o.f.  &  61.85/42\\
\noalign {\smallskip}
\hline
\noalign {\smallskip}
O VIII  & norm= 62$\pm$17 $\times$ 10$^{50}$ ph\,s$^{-1}$ \\ 
\noalign {\smallskip}
($\lambda$=18.97\AA)     & EW=2.2$\pm$0.6 \AA\\
\noalign {\smallskip}
N VII   & norm= 0$\pm$15 $\times$ 10$^{50}$ ph\,s$^{-1}$ \\
\noalign {\smallskip}
($\lambda$=24.78\AA)           & EW=0.0$\pm$ 0.6 \AA\\
\noalign {\smallskip}
C VI     & norm=118$\pm$41 $\times$ 10$^{50}$ ph\,s$^{-1}$ \\
\noalign {\smallskip}
($\lambda$=33.74\AA)        & EW=5.5$\pm$1.9 \AA\\
\noalign {\smallskip}
\noalign {\smallskip}
inclination    & 60$^{\rm o}$ $^{+5}_{-11}$\\
\noalign {\smallskip}
q                    & q$>$5.4\\
\noalign {\smallskip}
r$_{1}$  (in GM/c$^{2}$)                 & 1.64$^{+0.45}_{-0.14}$\\
\noalign {\smallskip}
r$_{2}$  (in GM/c$^{2}$)                   & 400 (fixed)\\
\noalign {\smallskip}
\hline
\noalign {\smallskip}
{\sc mbb}     &0.53$\pm$0.06 keV \\
\noalign {\smallskip}
            &  norm=0.58$\pm$0.17$\times$10$^{32}$\,m$^{0.5}$\\
\noalign {\smallskip}
\hline
\hline
\end{tabular}
\label{table:bestfit}
\end{table}

 To summarize, we find that the RGS spectrum 
 is consistent with an interpretation in
terms of weak relativistic lines. The formal significance of the O VIII 
and C VI lines is 3.6$\sigma$ and 2.9$\sigma$, respectively. 
  We do not exclude alternative spectral models, 
 but for example an interpretation
in terms of a dusty warm absorber giving deep edges at the wavelengths
of the blue edges of our broad lines is difficult given the lack of
deep narrow absorption lines in the spectrum (see also our tight limits
on the column density of any warm absorber in Fig.~4). 

\section{Summary and discussion}

 The present data analysis of the observation of ESO\,198-G24
 with XMM-Newton (January 24, 2001), has revealed that this 
Seyfert\,1 has an unabsorbed 2--10\,keV 
luminosity of about 1.1 $\times$10$^{-11}$\,erg\,s$^{-1}$.\\
 \indent We found both with EPIC and RGS 
that this source possesses a significantly steeper 
spectrum below $\sim$1.5--2\,keV than observed at higher X-ray energies, 
the so-called soft X-ray excess.   
 This soft X-ray emission seems to be a common feature in Seyfert 1 galaxies
observed up to now with {\sl XMM-Newton} (see Pounds \& Reeves \cite{Pounds2002}).
The shape of this soft excess emission 
is similar to those found in other high-luminosity Seyfert 1 galaxies.
(see for example  Fig.~1 in Pounds \& Reeves 2002). 
 The RGS spectra reveal no significant narrow absorption lines 
 suggesting that  if there is a warm absorber, 
it either has a relatively low column density, 
or a very high ionization  parameter.  
 The RGS data are well described by  the combination 
   a power-law,  a modified black body continuum,
 and weak relativistic soft X-ray lines from H-like ions 
\ion{O}{viii}, and \ion{C}{vi} Ly$_{\alpha}$, 
as may be observed with the RGS 
in at least two others Seyfert\,1 galaxies:  
 \object{Mrk 766}  and MCG-6-30-15
  (Branduardi-Raymont et al. \cite{BR2001}, 
Mason et al. \cite{Mason2003}). 
 However in our case the equivalent width of the relativistic lines is a factor
of 5--10 smaller than in the two sources mentioned above. 
  The presence of relativistic soft X-ray lines
 in AGN is a hotly debated issue
(e.g., Lee et al. \cite{Lee2001}, Branduardi-Raymont et al. \cite{BR2001}, 
Mason et al. \cite{Mason2003}). 
Here is not the place to discuss this issue in detail, but we note that
the strength of the proposed relativistic lines in our data are consistent
with those expected from ionized disk models, while the absence of strong
 absorption lines
makes the alternative dusty warm absorber less likely.
  However other interpretations are not definitely excluded.

 The EPIC data displayed a narrow emission line at 
about 6.4\,keV, corresponding to low or moderate ionized 
 iron (i.e., $\leq$ \ion{Fe}{xvii}).
 We find no significant broad (relativistic line) component, 
if any in our observation,  with only an upper limit of 75\,eV. 
Then the possible contribution from the inner part  
of an accretion disk where the gravitational effects of the BH are strong, 
is weak in the present {\sl XMM-Newton} observation. 
  We find for the narrow line a moderate equivalent width of about 60--70\,eV.
The narrow line at 6.4 keV seems to be an increasingly 
common feature in the higher quality spectral data emerging
from Chandra and XMM-Newton 
 (e.g., Mrk 509: Pounds et al. \cite{Pounds2001}, 
NGC\,5548: Yaqoob et al. \cite{Yaqoob2001}, 
\object{Mrk 205}: Reeves et al. \cite{Reeves2001}). 
 The emission medium responsible for the line component 
at 6.4\,keV can be find with search of possible response to 
  to continuum variation with high statistical data:  
 inner part of the nucleus (accretion disc), 
 BLR (as in NGC\,5548), 
 NLR, 
 and  molecular torus (as in Mrk 509 and  Mrk 205), etc...   
 The very weak ionized line associated to \ion{Fe}{xxvi}, if any, could be interpreted
in terms of fluorescence from H-like iron in the highly ionized inner disk material, 
or could also be the signature of a very ionized WA. 
While we show that the absorption edge found near 7.24\,keV cannot be produced
 by the WA.\\

\section*{ Acknowledgments }
This work is based on observations obtained with XMM-Newton, an ESA science mission 
with instruments and contributions directly funded by ESA Member States and the USA (NASA). 
 The authors would like to thank the anonymous referee for careful reading of the manuscript. 
D.P. acknowledges grant support from MPE fellowship (Germany).
The Space Research Organisation of the Netherlands (SRON)
is supported financially by NWO, the Netherlands Organisation for Scientific
Research. Support from a PPARC studentship is acknowledged by KLP.


\begin{thebibliography}{}
\bibitem[1989]{AG89} 
Anders, E. \& Grevesse, N., 1989, Geochimica et Cosmochimica Acta,  53, 197
\bibitem[1993]{A93} 
Antonucci, R.\ 1993, \araa,  31, 473 
\bibitem[2001]{Bal2001}
  Ballantyne, D.R. Iwasawa, K.,  Fabian, A.C., 2001, MNRAS, 323, 506 
\bibitem[1999]{Blackman99} 
Blackman, E.~G.  1999, MNRAS,  306, L25 
\bibitem[2001]{BR2001} 
Branduardi-Raymont, G., Sako, M., Kahn, S.~M., Brinkman, A.~C., Kaastra, J.~S., Page, M.~J., 2001, A\&A,  365, L140
 \bibitem[2002]{Colbert2002} 
Colbert, E.~J.~M., Weaver, K.~A., Krolik, J.~H., Mulchaey, J.~S., Mushotzky, R.~F.\ 2002, ApJ, 581, 182 
\bibitem[2001]{denHerder2001} 
den Herder, J.~W., Brinkman, A.~C., Kahn, S.~M., et al., 2001, A\&A,  365, L7 
\bibitem[1991]{George91} 
George, I.~M., Fabian, A.~C., 1991, MNRAS,  249, 352 
\bibitem[1994]{Ghisellini94} 
Ghisellini, G., Haardt, F., Matt, G., 1994, MNRAS,  267, 743 
\bibitem[2003]{Guainazzi2003}
Guainazzi, M. 2003, A\&A, 401, 903
\bibitem[2001]{Kaastra2001}
Kaastra, J.S., 2001, "X-ray Observations of AGN", in
Spectroscopic Challenges of Photoionized Plasmas, 
ASP Conference Series Vol. 247, p.39
\bibitem[2002a]{Kaastra2002a}
 Kaastra, J.S., Mewe, R., Raassen, A.J.J., 2002a, in  New Visions of the X-ray Universe in the XMM-Newton and Chandra Era,
  ed. F.A. Jansen, ESA, in press
\bibitem[2002b]{Kaastra2002b}
 Kaastra, J.~S., Steenbrugge, K.~C., Raassen, A.~J.~J., van der Meer, R.~L.~J., Brinkman A.~C., 
Liedahl, D.~A., Behar, E., de Rosa, A., 2002b, A\&A,  386, 427 
\bibitem[2001]{Kaspi2001} 
Kaspi, S., Brandt, W.~N., Netzer, H., et al., 2001, ApJ,  554, 216 
\bibitem[1987]{Krolik87} 
Krolik, J.~H. \& Kallman ,T.~R., 1987, ApJ,  320, L5 
\bibitem[1994]{Krolik94} 
Krolik, J.~H., Madau, P., Zycki, P.~T., 1994, ApJ,  420, L57 
\bibitem[1991]{Laor91} 
Laor, A., 1991, ApJ,  376, 90 
\bibitem[1993]{Leahy93} 
Leahy, D.~A. \& Creighton, J., 1993, MNRAS,  263, 314 
\bibitem[2001]{Lee2001} 
Lee, J.~C., Ogle, P.~M., Canizares, C.~R., Marshall, H.~L., 
Schulz, N.~S., Morales, R., Fabian, A.~C.,  Iwasawa, K.\ 2001, ApJ, 554, L13 
\bibitem[1995]{Magdziarz95} 
Magdziarz, P. \& Zdziarski ,A.~A., 1995, MNRAS,  273, 837 
\bibitem[1999]{Malizia99} 
Malizia, A., Bassani, L., Zhang, S.N., Dean, A.J., Paciesas, W.S., Palumbo, G.G.C. 1999, ApJ, 519, 637 
\bibitem[2003]{Mason2003} 
Mason, K.~O., Branduardi-Raymont, G., Ogle, P.~M., et al., 2003, ApJ,  582, 95 
\bibitem[1994]{Nandra94} 
Nandra, K. \& Pounds, K.~A., 1994, MNRAS,  268, 405. 
\bibitem[2001]{O'Brien2001} 
O'Brien, P.~T., Reeves, J.~N., Turner, M.~J.~L., et al., 2001, A\&A,  365, L122 
\bibitem[1997]{P97} 
Peterson, B.~M.\ 1997, An introduction to active galactic nuclei, New York Cambridge University Press.
\bibitem[1982]{Piccinotti82} 
Piccinotti, G., Mushotzky, R.~F., Boldt, E.~A., Holt, S.~S., Marshall, F.~E., 
Serlemitsos, P.~J., Shafer, R.~A., 1982, ApJ,  253, 485 
\bibitem[1990]{Pounds90} 
Pounds, K.~A., Nandra, K., Stewart, G.~C., 
George, I.~M., Fabian, A.~C., 1990, Nature,  344, 132. 
\bibitem[2001]{Pounds2001} 
Pounds, K., Reeves, J.N., O'Brien, P.T., Page, K., Turner, M., Nayakshin, S., 2001, ApJ,  559, 181 
\bibitem[2002]{Pounds2002} 
Pounds, K. \& Reeves, J., 2002, to appear 
in `New Visions of the X-ray Universe in the XMM-Newton and Chandra Era', 26-30 
November 2001, ESTEC, The Netherlands.,  [astro-ph/0201436]
\bibitem[2000]{Reeves2000} 
Reeves, J.~N. \& Turner, M.~J.~L., 2000, MNRAS,  316, 234 
\bibitem[2001]{Reeves2001} 
Reeves, J.~N., Turner, M.~J.~L., Pounds, K.~A., O'Brien, P.~T., Boller, T., Ferrando, P., 
Kendziorra, E., Vercellone, S., 2001, A\&A,  365, L134 
\bibitem[1999]{RB99} 
Rokaki, E.~\& Boisson, C.\ 1999, MNRAS, 307, 41 
\bibitem[1993]{Ross93}
Ross, R.R. \& Fabian, A.C., 1993, MNRAS, 261, 74
\bibitem[2001]{Strueder2001} 
Str{\" u}der, L., Briel, U., Dennerl, K., et al., 2001, A\&A,  365, L18 
\bibitem[1995]{Tanaka95} 
Tanaka, Y., Nandra, K., Fabian, A.~C., et al., 1995, Nature,  375, 659. 
\bibitem[1994]{Titarchuk94} 
Titarchuk, L., 1994, ApJ,  434, 570 
\bibitem[1993]{Turner93}
Turner, T.J., George, I.M., Mushotzky, R.F. 1993, ApJ, 412, 72 
\bibitem[2001]{Turner2001} 
Turner, M.~J.~L., Abbey, A., Arnaud, M., et al., 2001, A\&A,  365, L27 
\bibitem[2000]{Wilms2000} 
Wilms, J., Allen, A., McCray, R., 2000, ApJ,  542, 914
\bibitem[2001]{Wilms2001} 
Wilms, J., Reynolds, C.~S., Begelman, M.~C., Reeves, J., Molendi, S., Staubert, R.~;., Kendziorra E., 2001, MNRAS,  328, L27 
\bibitem[2001]{Yaqoob2001} 
Yaqoob, T., George, I.~M., Nandra, K., Turner, T.~J., Serlemitsos, P.~J., Mushotzky, R.~F., 2001, ApJ, 546, 759 
\end{thebibliography}
\end{document}